\documentclass[prd,twocolumn,showpacs,showkeys,floatfix]{revtex4}
\usepackage{amsmath,amssymb}
\usepackage{latexsym}
\usepackage{bm}

\usepackage{hyperref}
\usepackage[ansinew]{inputenc}
\makeatletter
\def\eqnarray{\stepcounter{equation}\let\@currentlabel=\theequation
\global\@eqnswtrue
\global\@eqcnt\z@\tabskip\@centering\let\\=\@eqncr
$$\halign to \displaywidth\bgroup\@eqnsel\hskip\@centering
  $\displaystyle\tabskip\z@{##}$&\global\@eqcnt\@ne
  \hfil${\;##\;}$\hfil
  &\global\@eqcnt\tw@ $\displaystyle\tabskip\z@{##}$\hfil
   \tabskip\@centering&\llap{##}\tabskip\z@\cr}
\makeatother

\begin{document}
\title{EXPANSION-FREE CAVITY EVOLUTION: SOME EXACT ANALYTICAL MODELS}
\author{A. Di Prisco}
\email{adiprisc@fisica.ciens.ucv.ve}
\affiliation{Escuela de F\'{\i}sica, Facultad de Ciencias,
Universidad Central de Venezuela, Caracas, Venezuela.}
\author{L. Herrera}
\email{laherrera@cantv.net.ve}
\affiliation{Departamento   de F\'{\i}sica Te\'orica e Historia de la F\'{\i}sica,
Universidad del Pa\'\i s Vasco, Bilbao, Spain}
\author{J. Ospino}\affiliation{Departamento de Matem\'atica Aplicada,  Universidad de Salamanca, Salamanca, Spain.}
\email{jhozcrae@usal.es}
\author{N. O. Santos}
\email{N.O.Santos@qmul.ac.uk}
\affiliation{School of Mathematical Sciences, Queen Mary
University of London, London E1 4NS, UK. and\\
 Observatoire de Paris, Universit\'e Pierre et Marie Curie,LERMA(ERGA) CNRS - UMR 8112, 94200 Ivry, France.}
\author{V. M. Vi\~na-Cervantes}
\email{vivi@ula.ve}
\affiliation{Centro de F\'{\i}sica Fundamental, Facultad de Ciencias, Universidad de Los Andes, M\'erida, Venezuela}
\begin{abstract}
We consider  spherically symmetric distributions of anisotropic fluids with a central vacuum cavity, evolving under the condition of vanishing expansion scalar. Some analytical solutions are found satisfying Darmois junction conditions on both delimiting  boundary surfaces, while some  others require the presence of thin shells on either (or both) boundary surfaces. The solutions here obtained model the evolution of  the vacuum cavity and the surrounding fluid distribution, emerging after a central explosion, thereby showing the potential of expansion--free condition for the study of that kind of problems. This study complements a previously published work where modeling of the evolution of such kind of systems  was achieved through a different kinematical condition.
\end{abstract}

\date{\today}
\maketitle
\section{INTRODUCTION}
Highly energetic explosions in self-gravitating fluid distributions are  common events in relativistic astrophysics (see for example \cite{1N,2N} and references therein). There are two main questions related to this issue:
\begin{itemize}
\item What is the energy source of  such  explosions?
\item How the system evolves after the explosion?
\end{itemize}

This work is devoted to the second of the above questions. Our approach is completely analytical, which presents both  advantages and disadvantages. On the one hand the analytical approach allows one to carry out a rather model independent  analysis, allowing in turn to obtain general conclusions about the pattern evolution of the system. On the other hand however, in order to  handle analytically the involved expressions, one needs to introduce simplifying assumptions which might exclude important physical phenomena.

We are interested in analytical models which even if might be relatively simple to analyze, still contain some of the essential features of a realistic situation.  It should be emphasized that we are not interested in the dynamics and the conditions of the creation of the cavity itself, but only in its evolution once  it is already formed.

A similar study to the one presented here was recently published \cite{3N}, however the main difference between the two being the kinematical condition imposed on the evolution of the system. In \cite{3N} it was assumed that  the proper radial distance between any two  infinitesimally close fluid elements remains constant all along the evolution (the purely areal evolution condition). Here we assume that the expansion scalar vanishes  for all fluid elements. As discussed in \cite {4N,5N,3N} either assumption creates conditions for  the existence of a vacuum cavity surrounding the centre. The dynamical instability of expansion-free fluids is studied in \cite{Herrera}.

We have two hypersurfaces delimiting the fluid. The external one separating the fluid distribution from a Schwarzschild  spacetime (for simplicity  we assume the evolution to be adiabatic) and the  internal one, delimiting the cavity within which we have Minkowski spacetime. It should be clear that for cavities with sizes of the order of 20 Mpc or smaller, the assumption of  a  spherically symmetric spacetime outside the cavity is quite reasonable, since the observed universe cannot be considered homogeneous on scales less than 150-300 Mpc. Nevertheless for  larger cavities it should be more appropriate to consider their embedding in an expanding Lema\^{\i}tre-Friedmann-Robertson-Walker spacetime (for the specific case of void modeling in expanding universes see \cite{6N,7N} and references therein).

Thus,  we have to consider junction conditions on both hypersurfaces. Depending on whether we impose Darmois conditions \cite{8N} or allow for the existence of thin shells \cite{9N}, different kinds of models are obtained. In this paper we are particularly interested in models satisfying Darmois conditions, however  some models presenting thin shells will be also exhibited.

The first known solution satisfying the vanishing expansion scalar  is due to Skripkin
\cite{13N,Mac}. It is worth noticing that Skripkin did not assume explicitly the expansion scalar to vanish, however the conditions he imposed (isotropic fluid with a constant energy-density) imply the vanishing of the expansion scalar. Unfortunately though, as it was shown in \cite{5N}, the Skripkin model does not satisfy Darmois junction conditions  (in principle Skripkin models with thin shells satisfying Israel conditions  can be  obtained).

It is our purpose here to find exact solutions under the vanishing expansion scalar condition and satisfying  Darmois junction conditions on both delimiting hypersurfaces.

For sake of generality we shall consider  an anisotropic  fluid. Indeed, as  it has been shown in \cite{4N}, in the case of isotropic  pressure, the expansion-free conditions imposes that the energy-density is independent on the time-like coordinate, which severely restricts the models (further arguments to justify such kind of fluid distributions may found in \cite{10N,11N,12N}  and references therein). A detailed description of  this kind of distribution, the Darmois junction conditions on both, the inner and the outer boundary surface,  as well as definitions of kinematical  variables, and basic equations, are given in section II.   The general description of  the expansion-free fluids is given in section III, whereas explicit solutions are found and described  in sections IV and V. Finally results are summarized in the last section.

\section{FLUID DISTRIBUTION, KINEMATICAL VARIABLES AND BASIC EQUATIONS}
We consider a spherically symmetric distribution  of collapsing
fluid, bounded by a spherical surface $\Sigma^{(e)}$. The fluid is
assumed to be locally anisotropic (principal stresses unequal).

Choosing comoving coordinates inside $\Sigma^{(e)}$, the general
interior metric can be written
\begin{equation}
ds^2_-=-A^2dt^2+B^2dr^2+R^2(d\theta^2+\sin^2\theta d\phi^2),
\label{1}
\end{equation}
where $A$, $B$ and $R$ are functions of $t$ and $r$ and are assumed
positive. We number the coordinates $x^0=t$, $x^1=r$, $x^2=\theta$
and $x^3=\phi$. Observe that $A$ and $B$ are dimensionless, whereas $R$ has the same dimension as $r$.

The matter energy-momentum $T_{\alpha\beta}^-$ inside $\Sigma^{(e)}$
has the form
\begin{equation}
T_{\alpha\beta}^-=(\mu +
P_{\perp})V_{\alpha}V_{\beta}+P_{\perp}g_{\alpha\beta}+\Pi \chi_{
\alpha}\chi_{\beta}, \label{3}
\end{equation}
where $\mu$ is the energy density, $\Pi\equiv P_r-P_\perp$, $P_r$ the radial pressure,
$P_{\perp}$ the tangential pressure, $V^{\alpha}$ the four velocity of the fluid,
and $\chi^{\alpha}$ a unit four vector along the radial direction. These quantities
satisfy
\begin{equation}
V^{\alpha}V_{\alpha}=-1,  \;\; \chi^{\alpha}\chi_{\alpha}=1, \;\chi^{\alpha}V_{\alpha}=0,
\end{equation}
since we assumed the metric (\ref{1}) comoving then
\begin{equation}
V^{\alpha}=A^{-1}\delta_0^{\alpha}, \;\;
\chi^{\alpha}=B^{-1}\delta^{\alpha}_1.\label{5}
\end{equation}

\subsection{Einstein equations}
From (\ref{1}) and (\ref{3}), Einstein equations
\begin{equation}
G_{\alpha \beta} = 8 \pi T_{\alpha \beta}, \label{Eeq}
\end{equation}
read
\begin{eqnarray}
8\pi T_{00}=8\pi \mu A^2
=\left(2\frac{\dot{B}}{B}+\frac{\dot{R}}{R}\right)\frac{\dot{R}}{R}\nonumber\\
-\left(\frac{A}{B}\right)^2\left[2\frac{R^{\prime\prime}}{R}+\left(\frac{R^{\prime}}{R}\right)^2
-2\frac{B^{\prime}}{B}\frac{R^{\prime}}{R}-\left(\frac{B}{R}\right)^2\right],
\label{12} \\
8\pi T_{01}=0 =-2\left(\frac{{\dot R}^{\prime}}{R} -\frac{\dot
B}{B}\frac{R^{\prime}}{R}-\frac{\dot
R}{R}\frac{A^{\prime}}{A}\right),
\label{13} \\
8\pi T_{11}=8\pi P_rB^2 \nonumber\\
=-\left(\frac{B}{A}\right)^2\left[2\frac{\ddot{R}}{R}
-\left(2\frac{\dot A}{A}-\frac{\dot{R}}{R}\right)\frac{\dot R}{R}\right]\nonumber\\
+\left(2\frac{A^{\prime}}{A}+\frac{R^{\prime}}{R}\right)\frac{R^{\prime}}{R}-\left(\frac{B}{R}\right)^2,
\label{14} \\
8\pi T_{22}=\frac{8\pi}{\sin^2\theta}T_{33}=8\pi P_{\perp}R^2\nonumber \\
=-\left(\frac{R}{A}\right)^2\left[\frac{\ddot{B}}{B}+\frac{\ddot{R}}{R}
-\frac{\dot{A}}{A}\left(\frac{\dot{B}}{B}+\frac{\dot{R}}{R}\right)
+\frac{\dot{B}}{B}\frac{\dot{R}}{R}\right]\nonumber\\
+\left(\frac{R}{B}\right)^2\left[\frac{A^{\prime\prime}}{A}
+\frac{R^{\prime\prime}}{R}-\frac{A^{\prime}}{A}\frac{B^{\prime}}{B}
+\left(\frac{A^{\prime}}{A}-\frac{B^{\prime}}{B}\right)\frac{R^{\prime}}{R}\right],\label{15}
\end{eqnarray}
where the  prime stands for $r$
differentiation and the dot stands for differentiation with respect to $t$.

\subsection{Kinematical variables and the mass function}
The four acceleration   $a_{\alpha}$ is given by
\begin{equation}
a_{\alpha}=V_{\alpha ;\beta}V^{\beta}, \label{4b}
\end{equation}
from (\ref{4b}) and (\ref{5}) we obtain
\begin{equation}
a_1=\frac{A^{\prime}}{A}, \;\;
a^2=a^{\alpha}a_{\alpha}=\left(\frac{A^{\prime}}{AB}\right)^2
\label{5c}
\end{equation}
with  $a^\alpha= a \chi^\alpha$. The expansion $\Theta$ is given by
\begin{equation}
\Theta={V^{\alpha}}_{;\alpha}=\frac{1}{A}\left(\frac{\dot{B}}{B}+2\frac{\dot{R}}{R}\right).\label{th}
\end{equation}
and for the shear we have
\begin{equation}
\sigma_{\alpha\beta}=V_{(\alpha
;\beta)}+a_{(\alpha}V_{\beta)}-\frac{1}{3}\Theta h_{\alpha\beta}.
\label{4a}
\end{equation}
where $h_{\alpha \beta} = g_{\alpha \beta} + V_\alpha V_\beta $.
Using (\ref{5}) we obtain the non-vanishing components of
(\ref{4a})
\begin{equation}
\sigma_{11}=\frac{2}{3}B^2\sigma, \;\;
\sigma_{22}=\frac{\sigma_{33}}{\sin^2\theta}=-\frac{1}{3}R^2\sigma,
 \label{5a}
\end{equation}
with
\begin{equation}
\sigma^{\alpha\beta}\sigma_{\alpha\beta}=\frac{2}{3}\sigma^2,
\label{5b}
\end{equation}
being
\begin{equation}
\sigma=\frac{1}{A}\left(\frac{\dot{B}}{B}-\frac{\dot{R}}{R}\right).\label{5b1}
\end{equation}
$\sigma_{\alpha\beta}$ may also be written as
\begin{equation}
\sigma_{\alpha \beta}= \sigma \left(\chi_\alpha \chi_\beta -
\frac{1}{3} h_{\alpha \beta}\right). \label{sh}
\end{equation}

Next, the mass function $m(t,r)$ introduced by Misner and Sharp \cite{14N} (see also \cite{15N}) is
given by
\begin{equation}
m=\frac{R^3}{2}{R_{23}}^{23} =\frac{R}{2}
\left[\left(\frac{\dot{R}}{A}\right)^2
-\left(\frac{R^{\prime}}{B}\right)^2+1\right].
 \label{18}
\end{equation}

We can define the velocity $U$ of the collapsing
fluid as the variation of the areal  radius $R$ with respect to proper time, i.e.
\begin{equation}
U=\frac{\dot R}{A}. \label{19}
\end{equation}
Then (\ref{18}) can be rewritten as
\begin{equation}
E \equiv \frac{R^{\prime}}{B}=\left[1+U^2-2\frac{m(t,r)}{R}\right]^{1/2}.
\label{20}
\end{equation}
With the above  we can express (\ref{13}) as
\begin{equation}
0=\frac{1}{3 R^{\prime}}(\Theta-\sigma)^{\prime}
-\frac{\sigma}{R}.\label{21a}
\end{equation}
While from (\ref{18}) we have
\begin{eqnarray}
\dot m=-4\pi P_r \dot RR^2,
\label{22}
\end{eqnarray}
and
\begin{eqnarray}
m^{\prime}=4\pi\mu R^{\prime}R^2.
\label{27}
\end{eqnarray}
Combining (\ref{22}) and (\ref{27})  we obtain
\begin{equation}
\dot \mu R^\prime +P^\prime _r \dot R+(P_r+\mu)\left(\dot R^\prime
+2R^\prime \frac{\dot R}{R}\right)=0.\label{j}
\end{equation}

The integration of (\ref{27}) produces
\begin{equation}
m=\int^{r}_{0}4\pi\mu R^2R^{\prime}dr\label{27int}
\end{equation}
where a regular centre to the distribution, $m(0)=0$ is assumed. We may partially integrate (\ref{27int}) to obtain
\begin{equation}
3\frac{m}{R^3} = 4\pi \mu - \frac{4\pi}{R^3} \int^r_0\mu^{\prime}R^3 dr.
\label{3m/R3}
\end{equation}

\subsection{Junction conditions}
Outside $\Sigma^{(e)}$ (whose equation is $r=r_e=$ constant) we assume the Schwarzschild
spacetime, described by
\begin{equation}
ds^2=-\left(1-2\frac{M}{\rho}\right)dv^2-2d\rho dv+\rho^2(d\theta^2
+\sin^2\theta
d\phi^2) \label{1int},
\end{equation}
where $M$  denotes the total mass and  $v$ is the retarded time.

The matching of the  adiabatic fluid  sphere  to
Schwarzschild
spacetime, on the surface $r=r_e=$ constant (or $\rho=\rho(v)_e$ in the coordinates  of (\ref{1int})), requires  the continuity of the first and second differential forms (Darmois conditions), implying (see \cite{16N} for details)
\begin{eqnarray}
 Adt\stackrel{\Sigma^{(e)}}{=}dv \left(1-2\frac{M}{\rho}\right), \label{junction1fn}\\
R\stackrel{\Sigma^{(e)}}{=}\rho(v), \label{junction1f2n}\\
m(t,r)\stackrel{\Sigma^{(e)}}{=}M, \label{junction1}
\end{eqnarray}
and
\begin{widetext}
\begin{eqnarray}
2\left(\frac{{\dot R}^{\prime}}{R}-\frac{\dot B}{B}\frac{R^{\prime}}{R}-\frac{\dot R}{R}\frac{A^{\prime}}{A}\right)
\stackrel{\Sigma^{(e)}}{=}-\frac{B}{A}\left[2\frac{\ddot R}{R}
-\left(2\frac{\dot A}{A}
-\frac{\dot R}{R}\right)\frac{\dot R}{R}\right]+\frac{A}{B}\left[\left(2\frac{A^{\prime}}{A}
+\frac{R^{\prime}}{R}\right)\frac{R^{\prime}}{R}-\left(\frac{B}{R}\right)^2\right],
\label{j2}
\end{eqnarray}
\end{widetext}
where $\stackrel{\Sigma^{(e)}}{=}$ means that both sides of the equation
are evaluated on $\Sigma^{(e)}$ (observe a misprint in equation (40) in \cite{16N} and a slight difference in notation).

Comparing (\ref{j2}) with  (\ref{13}) and (\ref{14}) one obtains
\begin{equation}
P_r\stackrel{\Sigma^{(e)}}{=}0.\label{j3}
\end{equation}

As we mentioned in the introduction, the expansion-free models  present an internal vacuum cavity. If we call $\Sigma^{(i)} $ the boundary surface between the cavity and the fluid, described by the equation $r=r_i=$ constant, then the matching of the Minkowski spacetime within the cavity to the fluid distribution, implies
\begin{eqnarray}
m(t,r)\stackrel{\Sigma^{(i)}}{=}0, \label{junction1i}\\
P_r\stackrel{\Sigma^{(i)}}{=}0.\label{j3i}
\end{eqnarray}

\subsection{Weyl tensor}
The Weyl tensor is defined through the  Riemann tensor
$R^{\rho}_{\alpha \beta \mu}$, the  Ricci tensor
$R_{\alpha\beta}$ and the curvature scalar $\cal R$, as
\begin{eqnarray}
C^{\rho}_{\alpha \beta \mu}=R^\rho_{\alpha \beta \mu}-\frac{1}{2}
R^\rho_{\beta}g_{\alpha \mu}+\frac{1}{2}R_{\alpha \beta}\delta
^\rho_{\mu}-\frac{1}{2}R_{\alpha \mu}\delta^\rho_\beta, \nonumber\\
+\frac{1}{2}R^\rho_\mu g_{\alpha \beta}+\frac{1}{6}{\cal
R}(\delta^\rho_\beta g_{\alpha \mu}-g_{\alpha
\beta}\delta^\rho_\mu). \label{34}
\end{eqnarray}
The electric  part of  Weyl tensor is defined by (the magnetic part vanishes due to the spherical symmetry)
\begin{equation}
E_{\alpha \beta} = C_{\alpha \mu \beta \nu} V^\mu V^\nu,
\label{elec}
\end{equation}
with the following non-vanishing components
\begin{eqnarray}
E_{11}=\frac{2}{3}B^2 {\cal E},\nonumber \\
E_{22}=-\frac{1}{3} R^2 {\cal E}, \nonumber \\
E_{33}= E_{22} \sin^2{\theta}, \label{ecomp}
\end{eqnarray}
where

\begin{eqnarray}
&&{\cal E}= \frac{1}{2 A^2}\left[\frac{\ddot R}{R} - \frac{\ddot B}{B} - \left(\frac{\dot R}{R} - \frac{\dot B}{B}\right)\left(\frac{\dot A}{A} + \frac{\dot R}{R}\right)\right]\\
&+& \frac{1}{2 B^2} \left[\frac{A^{\prime\prime}}{A} - \frac{R^{\prime\prime}}{R} + \left(\frac{B^{\prime}}{B} + \frac{R^{\prime}}{R}\right)\left(\frac{R^{\prime}}{R}-\frac{A^{\prime}}{A}\right)\right] - \frac{1}{2 R^2}.\nonumber
\label{E}
\end{eqnarray}
Observe that we may also write $E_{\alpha\beta}$ as
\begin{equation}
E_{\alpha \beta}={\cal E}\left(\chi_\alpha
\chi_\beta-\frac{1}{3}h_{\alpha \beta}\right). \label{52}
\end{equation}

Using (\ref{12}), (\ref{14}), (\ref{15}) with (\ref{18}) and (\ref{E}) we obtain
\begin{equation}
3\frac{m}{R^3}=4\pi (\mu-\Pi) - \cal{E}.
\label{mE}
\end{equation}
Finally, comparing (\ref{3m/R3}) with (\ref{mE}) we get
\begin{equation}
{\cal E} = \frac{4\pi}{R^3} \int^r_0\mu^{\prime}R^3 dr-4\pi \Pi,
\label{Emat}
\end{equation}
which exhibits the  dependence of the Weyl tensor on  energy-density inhomogeneity and local anisotropy of pressure.

\subsection{ Bianchi identities and an evolution equation  for the Weyl tensor}
The two independent components of Bianchi identities for the system under consideration read (see \cite{4N} for details):
\begin{equation}
\dot{\mu}+( \mu + P_r)A\Theta-2\Pi \frac{\dot
R}{R}=0,\label{a}
\end{equation}

\begin{equation}
P_r ^\prime+(\mu +
P_r)\frac{A^\prime}{A}+2\Pi\frac{R^\prime}{R}=0.\label{b}
\end{equation}

Finally, the following evolution equation for the Weyl tensor will be used, which may be derived from the Bianchi identities \cite{17N,18N} (see also \cite{11N} for details)
\begin{equation}
\left[{\cal E}-4\pi( \mu-\Pi) \right]^{\dot{}}+3\left[{\cal E}-4\pi(
\mu+P_{\perp})\right]\frac{\dot R}{R}=0. \label{g}
\end{equation}

\section{THE EXPANSION-FREE FLUID}
If we assume the evolution to be expansion-free, i.e. $\Theta=0$,
then from (\ref{th}) we have
\begin{equation}
\frac{\dot B}{B}=-2\frac{\dot R}{R}\,\, \Rightarrow \,\,
B=\frac{h}{R^2},\label{th1}
\end{equation}
and after substituting (\ref{th1}) into (\ref{13}), we get
\begin{equation}
A=\frac{\dot R R^2}{\tau},\label{a1}
\end{equation}
where $\tau (t)$  and $h(r)$ are arbitrary functions of their arguments. Thus, without loss of generality,  we shall take $h=1$ (observe that with this choice  a unit constant with dimensions $[r^4]$ is assumed to multiply $dr^2$).

Next, using the equations  (\ref{22}), (\ref{27}), (\ref{a}) and
(\ref{th1}), the physical variables $\mu$, $P_r$, and $\Pi$, can
be written through the mass function $m$ and the metric function $R$
as follows,
\begin{eqnarray}
4\pi \mu=\frac{m^\prime}{R^\prime R^2}, \label{mug}\\
4\pi P_r=-\frac{\dot m}{\dot R R^2},\label{pg}\\
\Pi=\frac{\dot \mu R}{2 \dot R}. \label{ppg}
\end{eqnarray}
On the other hand, combining (\ref{18}), (\ref{th1}) and
(\ref{a1}) we obtain,
\begin{equation}
m(t,r)=\frac{1}{2}\left(\frac{\tau ^2}{R^3}-R^{\prime 2}R^5+R\right). \label{mg}
\end{equation}
To completely determine  the metric we need to give an extra
condition. In the next sections we provide some options.

\section{SOLUTION I}
The first family of solutions will be obtained by choosing  the function $m(t,r)$ as follows \cite{19N},
\begin{equation}
2m(t,r)=jR+\frac{1}{3}kR^3+\frac{1}{5}lR^5, \label{mM}
\end{equation}
where $j(t)$, $k(t)$ and $l(t)$ are so far arbitrary functions of $t$.
Substituting (\ref{mM}) into  (\ref{mug}) we
obtain  for the energy density
\begin{equation}
8\pi \mu =\frac{j}{R^2}+k+lR^2. \label{muM}
\end{equation}
\noindent Then, taking the $r$ derivative of (\ref{mg}) and using
(\ref{mug}) and (\ref{muM}) we obtain,
\begin{equation}
j-1+kR^2+lR^4+3\frac{\tau^2}{R^4}+2 R^{\prime\prime}R^5+5R^{\prime 2}R^4=0.\label{eq0}
\end{equation}
Combining (\ref{mg}), (\ref{mM}) and
(\ref{eq0}), we get
\begin{equation}
4(j-1)+2kR^2+\frac{8}{5}lR^4+2R^{\prime\prime}R^5+8R^{\prime 2}R^4 =0,\label{eq1}
\end{equation}
which by introducing the  new function $Z\equiv R^2$, can be transformed into
\begin{equation}
a+bZ+cZ^2+Z^{\prime\prime}Z^2+\frac{3}{2}Z^{\prime 2}Z=0,\label{eq2}
\end{equation}
where
\begin{equation}
a(t)\equiv 4(j-1),\quad b(t)\equiv 2k, \quad c(t)\equiv \frac{8}{5}l. \label{const}
\end{equation}
Finally, feeding back  $Y(Z)=Z^{\prime 2}$ into (\ref{eq2}),
leads to the first order equation
\begin{equation}
\frac{dY}{dZ}+3\frac{Y}{Z}=-2\left(\frac{a}{Z^2}+\frac{b}{Z}+c\right),\label{eq3}
\end{equation}
and after integration we  obtain
\begin{equation}
Y(Z)=Z^{\prime 2}
=-\left(\frac{a}{Z}+\frac{2}{3}b+\frac{1}{2}cZ\right).\label{eq4}
\end{equation}
This last  equation  can be integrated in its general
form, however in order to obtain simple models, in what follows we shall consider specific cases.

\subsection{a=0}
For $a=0$ ($j=1$), integration of
(\ref{eq4})produces,
\begin{equation}
Z=-\frac{4}{3}\frac{b}{c}-\frac{1}{8}c(r+\beta)^2,
\end{equation}
implying
\begin{equation}
R=\left[-\frac{4}{3}\frac{b}{c}-\frac{1}{8}c(r+\beta)^2\right]^{1/2}=\left[-\frac{5k}{3l}-\frac{l}{5}(r+\beta)^2\right]^{1/2},
\label{eq5}
\end{equation}
where $\beta (t)$ is an arbitrary function of t.
Then, using  (\ref{muM}), (\ref{pg}) and
(\ref{ppg}) the physical variables, $\mu$, $P_r$ and $\Pi$, can be
expressed as
\begin{equation}
8\pi
\mu=\frac{1}{R^2}+k+lR^2,\label{muM1}
\end{equation}
\begin{equation}
8\pi P_r=-\frac{1+kR^2+lR^4}{R^2}-\frac{R}{\dot R}\left(\frac{\dot k}{3}+\frac{\dot l}{5}R^2\right),\label{pM1}\end{equation}
\begin{equation}
8\pi
\Pi=-\frac{1}{R^2}+R^2\left(l+\frac{\dot k+\dot l R^2}{2R\dot R}\right).\label{ppM1}
\end{equation}

This subfamily of solutions may satisfy Darmois conditions. Indeed junctions conditions (\ref{junction1}), (\ref{j3}) and (\ref{junction1i}), (\ref{j3i}) lead to two independent equations for the three functions $b(t)$, $c(t)$ and $\beta(t)$ which can be satisfied by any convenient choice of one of these functions. However for all the attempts we have carried out resulting expressions were extremely cumbersome and therefore we shall not pursue this direction further.

\subsection{b=c=0}
For $b(t)=c(t)=0$, integration of  (\ref{eq4}) produces
\begin{equation}
Z=\left(\frac{3}{2}\sqrt{-a}r+\beta\right)^{2/3},
\end{equation}
or
\begin{equation}
R=\left(\frac{3}{2}\sqrt{-a}r+\beta\right)^{1/3}.\label{eq6}
\end{equation}
Then, using  (\ref{muM}), (\ref{pg}) and
(\ref{ppg}) the physical variables $\mu$, $P_r$ and $\Pi$ read,
\begin{eqnarray}
8\pi\mu
=j\left(\frac{3}{2}\sqrt{-a}r+\beta\right)^{-2/3},\label{muM2}\\
8\pi P_r=2{\dot j}\left(\frac{3}{2}\sqrt{-a}r
+\beta\right)^{1/3}\left(\frac{1}{2}\frac{\dot a}{\sqrt{-a}}r
-\frac{2}{3}{\dot\beta}\right)^{-1} \nonumber\\
-j\left(\frac{3}{2}\sqrt{-a}r+\beta\right)^{-2/3},\label{pM2}\\
8\pi \Pi=-{\dot j}\left(\frac{3}{2}\sqrt{-a}r+\beta\right)^{1/3}
\left(\frac{1}{2}\frac{\dot a}{\sqrt{-a}}r-\frac{2}{3}{\dot\beta}\right)^{-1} \nonumber\\
-j\left(\frac{3}{2}\sqrt{-a}r+\beta\right)^{-2/3}.\label{ppM2}
\end{eqnarray}

It is a simple matter to check that this subfamily of solution does not satisfy Darmois conditions  on the internal boundary surface $r=r_i$. Accordingly a thin shell should be present there.

\section{SOLUTION II}
The next  family of solutions  is found with the additional assumption $P_r=0$. Fluid spheres with tangential stresses alone, have a long and a venerable history starting with  Lema\^{\i}tre seminal paper \cite{20N}. Afterwards, many authors have used such kind of fluid distribution for different purposes (see for example \cite{21N, 22N, 23N, 24N, 25N, 26N, 27N} and references therein).

Thus,  assuming  $P_r=0$, we find   from (\ref{j})
\begin{equation}
\mu=\frac{C_1}{R^\prime R^2}\,\, \Rightarrow\,\, R^3=3\int
\frac{C_1}{\mu}dr+C_2,\label{01}
\end{equation}
where $C_1(r)$ and $C_2(t)$ are two functions of
integration. Using  (\ref{mug}) in (\ref{01}), the former can be easily identified as
\begin{equation}
C_1=\frac{m'}{4\pi}.
\label{c1}
\end{equation}

 Now, feeding back (\ref{th1}) in  (\ref{a}) we obtain
for the tangential pressure $P_\perp$
\begin{equation}
P_\perp=-\frac{\dot \mu}{2}\frac{R}{\dot R}.\label{pmo}
\end{equation}

In what follows we shall impose different additional  restrictions in order to obtain different subfamilies of  models.

\subsection{$P_{\perp}=\alpha \mu$}
Assuming
\begin{equation}
P_r=0, \;\; P_{\perp}=\alpha\mu, \label{N3}
\end{equation}
where $\alpha$ is a constant, then (\ref{b}) becomes
\begin{equation}
\frac{{\dot R}^{\prime}}{\dot R}=2(\alpha -1)\frac{R^{\prime}}{R}, \label{N4}
\end{equation}
and after integration
\begin{equation}
{\dot R}=fR^{2(\alpha-1)}, \;\; R^{\prime}=gR^{2(\alpha-1)}. \label{N5}
\end{equation}
with $f(t)$ and $g(r)$ denoting arbitrary functions of their arguments. Then from (\ref{N5}) we obtain,
\begin{equation}
R^{3-2\alpha}=\psi(t)+\chi(r), \label{N6}
\end{equation}
with
\begin{equation}
\psi(t)=(3-2\alpha)\int fdt, \;\; \chi(r)=(3-2\alpha)\int gdr. \label{7}
\end{equation}
Without loss of generality we may choose $\tau(t)=f$, implying because of (\ref{a1}) and (\ref{N5}),
\begin{equation}
A=R^{2\alpha}. \label{N8}
\end{equation}
Considering (\ref{N3}) with (\ref{22}) we have
\begin{equation}
m=m(r), \label{N9}
\end{equation}
and from (\ref{18}) with (\ref{th1}) and (\ref{N8}) we obtain
\begin{equation}
m(r)=\frac{R}{2}\left(\frac{{\dot R}^2}{R^{4\alpha}}-g^2R^{4\alpha}+1\right). \label{N10}
\end{equation}
Matching conditions (\ref{junction1}) and (\ref{junction1i}) on  the hypersurfaces $r=r_e$  and $r=r_i$,  together with    (\ref{N10}), produce
\begin{equation}
{\dot R}^2\stackrel{\Sigma^{(i)}}{=}R^{4\alpha}(g^2R^{4\alpha}-1), \label{N12}
\end{equation}
and
\begin{equation}
{\dot R}^2\stackrel{\Sigma^{(e)}}{=}R^{4\alpha}\left(\frac{2M}{R}+g^2R^{4\alpha}-1\right). \label{N12b}
\end{equation}
Equation (\ref{N12}) can be integrated  for any given value of $\alpha$ (the particular case of dust, $\alpha=0$ has already been considered in \cite{5N}).

For  $\alpha=1/2$ the solution reads
\begin{equation}
R\stackrel{\Sigma^{(i)}}{=}\frac{1}{g\cos(t+t_0)}, \label{N15}
\end{equation}
and from (\ref{N6}) evaluated on $\Sigma^{(i)}$,
\begin{equation}
\psi(t)\stackrel{\Sigma^{(i)}}{=}\frac{1}{g^2\cos^2(t+t_0)}+\chi. \label{N16}
\end{equation}

Thus the time dependence of all variables is fully determined. The radial dependence  ($C_1$ or $\chi$) can be obtained from the initial data.

However, this solution is inconsistent for any $\alpha\geq 0$ with junction conditions satisfied on both hypersurfaces.
Indeed, from (\ref{N12}) and   (\ref{N12b}) evaluated at $t=0$ we get
\begin{equation}
R_i^{4\alpha}(0)=\frac{1}{2g^2}\left[1+\left(1+4g^2\dot R_i^2(0)\right)^{1/2}\right],
\label{n1}
\end{equation}
and
\begin{eqnarray}
R_e^{4\alpha}(0)=\frac{1}{2g^2} \nonumber\\
\times\left\{1-\frac{2M}{R_e(0)}
+\left[\left(1-\frac{2M}{R_e(0)}\right)^2+4g^2\dot R_e^2(0)\right]^{1/2}\right\}.
\label{n2}
\end{eqnarray}
Now substracting (\ref{n2}) from (\ref{n1}) we obtain
\begin{widetext}
\begin{equation}
R_i^{4\alpha}(0)-R_e^{4\alpha}(0)=\frac{1}{2g^2}
\left[1+\left(1+4g^2{\dot R}_i^2(0)\right)^{1/2}\right] -\frac{1}{2g^2}\left\{1-\frac{2M}{R_e(0)}+\left[\left(1-\frac{2M}{R_e(0)}\right)^2
+4g^2{\dot R}_e^2(0)\right]^{1/2}\right\},
\label{n3}
\end{equation}
\end{widetext}
which should be  negative for all $t$ (if $\alpha>0$) (or zero if $\alpha=0$) since obviously $R_e(t)>R_i(t)$. Unfortunately this is not the case.

In fact, from the definition of $U$, using (\ref{a1}) and (\ref{N8}) we obtain
\begin{equation}
U_i=\frac{f}{R_i^2}=\frac{\dot R_i}{R_i^{2\alpha}},
\label{n4}
\end{equation}
and
\begin{equation}
U_e=\frac{f}{R_e^2}=\frac{\dot R_e}{R_e^{2\alpha}}.
\label{n5}
\end{equation}
Since $\alpha<1$ and $R_e>R_i$ it follows from the above that
\begin{equation}
\dot R_i>\dot R_e.
\label{n6}
\end{equation}
Then using (\ref{n6}) in (\ref{n3}) we obtain at once that $R_i^{4\alpha}(0)>R_e^{4\alpha}(0)$ which is unacceptable.

Therefore such models require the presence of thin shells at either (or both) boundary surfaces.

\subsection{$\mu =\mu _0 C_1/r^2$}
Next, the following subfamily of solutions will be obtained under the assumption  that the energy-density is separable. Thus we write
\begin{equation}
\mu = \frac{\mu _0 C_1}{r^2},\label{mumo}
\end{equation}
where $\mu_0(t)$  is an  arbitrary function of its argument and $C_1(r)$ is given by (\ref{c1}).
Then substituting (\ref{mumo}) into (\ref{01}) and
(\ref{pmo}), we get
\begin{eqnarray}
R= \left(\frac{r^3}{\mu _0}+C_2\right)^{1/3}
\label{1tex}
\end{eqnarray}
\begin{eqnarray}
P_\perp=\frac{3}{2}\frac{\mu _0 \dot \mu _0 C_1}{r^2}\left(\frac{r^3+\mu_0C_2}
{{\dot\mu}_0 r^3-\mu _0^2 {\dot C}_2}\right),\label{1texbis}
\end{eqnarray}
and from (\ref{mg}) and (\ref{1tex}) we obtain
\begin{equation}
m=\frac{1}{2}\left[\frac{\tau^2}{R^3}-R\left(\frac{r^4}{\mu_0^2}-1\right)\right].
\label{op21}
\end{equation}
Evaluating this expression on $\Sigma^{(i)}$ by using (\ref{junction1i}) we obtain
\begin{equation}
R^4_i=\tau^2\left(\frac{r^4_i}{\mu_0^2}-1\right)^{-1},
\label{0p22}
\end{equation}
implying the restriction $r^4_i> \mu_0^2$ for all $t$.

Next, from the definition of $U$ and (\ref{a1}) we get
\begin{equation}
U=\frac{\tau}{R^2},
\label{0p23}
\end{equation}
implying
\begin{equation}
U^2_i=\frac{r^4_i}{\mu_0^2}-1.
\label{0p24}
\end{equation}
Thus, from the absence of superluminal velocities ($U<1$) and  (\ref{0p24}) we need to impose
\begin{equation}
\sqrt{2}\mu_0>r^2_i>\mu_0.
\label{0p25}
\end{equation}
Hence from (\ref{0p23}) we have
\begin{equation}
U_e=U_i\left(\frac{R_i}{R_e}\right)^2,
\label{0p26}
\end{equation}
and since $R_e>R_i$ then from (\ref{0p26}) it follows that the inner boundary surface moves faster than the outer one.

Now, since $\tau$ is a function of $t$ which may be chosen arbitrarily, we put without loss of generality
\begin{equation}
\tau=R^2_i\dot R_i,\label{0p27}
\end{equation}
implying
\begin{equation}
A_i=1, \;\;   \dot R_i=U_i, \;\;  \tau=R^2_iU_i=U_e R^2_e.
\label{0p28}
\end{equation}

Let us now consider the junction conditions at the outer boundary surface $\Sigma^{(e)}$. Using (\ref{junction1}), (\ref{op21}) and (\ref{0p28}) we obtain
\begin{equation}
R_e=2M\left(U^2_e-\frac{r^4_e}{\mu_0}+1\right)^{-1},\label{0p29}
\end{equation}
imposing the restriction on $r_e$,
\begin{equation}
\sqrt{2} \mu_0>r^2_e>\mu_0,\label{0p30}
\end{equation}
which is the same as imposed on $r_i$.

Thus if we provide the time dependence of $\mu_0$ then (\ref{0p22}) allows  to determine  the time dependence of $R_i$, and using this result in (\ref{1tex}) we obtain $C_2(t)$. Finally evaluating (\ref{1tex}) on  $\Sigma^{(e)}$ we get $R_e$.

Alternatively we may provide the time dependence of, either $R_i$ or $R_e$ which allows also to find the time dependence of all variables. We shall next  exhibit  a specific model obtained from the additional restriction of conformal flatness.

\subsubsection{Conformally flat models}
Instead of assuming any specific time dependence for any of the variables considered above, we shall assume here that the fluid is conformally flat, i.e.  we shall assume ${\cal E}=0$ then we obtain from (\ref{mE}) (with $P_r=0$)
 \begin{equation}
\mu+P_\perp=\frac{3m}{4 \pi R^3},\label{0p31}
\end{equation}
Next, evaluating (\ref{0p31}) at  $\Sigma^{(i)}$, using (\ref{pmo}) and (\ref{mumo}), we obtain
\begin{equation}
R_i^2=\alpha \mu_0,\label{0p33}
\end{equation}
where $\alpha$ is a constant.
Feeding back (\ref{0p33}) into (\ref{0p24}) we obtain
\begin{equation}
\dot R_i^2=\left(\frac{\beta}{R_i^2}\right)^2-1, \label{0p34}
\end{equation}
with $\beta^2\equiv r^4_i \alpha^2$.
Thus
\begin{equation}
\dot R_i=\pm \left[\left(\frac{\beta}{R_i^2}\right)^2-1\right]^{1/2}, \label{0p35}
\end{equation}
where $+$ ($-$) corresponds to expanding (contracting) configurations.
The integral of (\ref{0p35}) is expressed through elliptic functions, and we shall not display explicitly the solution.

The time dependence of the remaining functions is obtained as in the previous case.

Finally, we need to determine $C_1(r)$, for doing that we may proceed as follows:
from (\ref{27}) and (\ref{mumo}) we have
\begin{equation}
m'=\frac{4\pi}{r^2}R'R^2\mu_0C_1.
\label{f}
\end{equation}
Taking the $r$ derivative of $m$ from (\ref{op21}), feeding it back into (\ref{f}) and evaluating the result at $t=0$ we obtain $C_1(r)$.

These solutions are regular everywhere (within the fluid distribution) and satisfy basic physical restrictions such as positivity of energy density distribution and $\mu>P_{\perp}$.

\section{CONCLUSIONS}
We have revisited the original Skripkin problem, but now allowing for a more general fluid distribution. This enabled us to find analytical models which satisfy Darmois junction conditions on both delimiting boundary surfaces.
Obviously, relaxing Darmois conditions on either (or both) boundary hypesurfaces enlarges the family of possible solutions, some of them are explicitly exhibited in previous sections (for models of voids within the thin wall approximation, see  \cite{v4}-\cite{v6} and references therein).

The most important contribution in this manuscript  is to illustrate the advantages of using the expansionfree condition for the modeling of cavity evolution. For that purpose we have integrated analytically the pertinent equations under a variety of circumstances, proving that such an integration could be achieved  without undue difficulty. The obtained solutions were not intended to describe any specific astrophysical scenario but just to bring out  the potential of the expansion--free condition  to model situation where cavities are expected to appear.

Possibly our solutions could be applied as toy models of localized  systems such as supernova explosion models. Most likely the combination of the vanishing expansion scalar condition with numerical integration of  corresponding equations would produce more physically meaningful models.

It is worth mentioning the fact that in spite of the vanishing of the expansion scalar, the energy-density changes with time. This effect is due to the work done by the anisotropic force as it appears from (\ref{a}).

\begin{acknowledgments}
LH wishes to thank Fundaci\'on Empresas Polar for financial support. ADP  acknowledges hospitality of the Departamento de F\'isica Te\'orica e Historia de la F\'\i sica at Universidad del Pa\'\i s Vasco. JO wishes to thank Universidad de Salamanca  
for financial support under   grant  No. FK10.
\end{acknowledgments}

\end{document}